\documentclass[11pt]{article}
\usepackage{amsmath,amsfonts,amssymb} % ,epsfig}
\usepackage{xspace}

\newcommand{\SLSP}{\textsc{slsp}\xspace}
\newcommand{\HSP}{\textsc{hsp}\xspace}

\newcommand{\Z}{{\mathbb Z}}

\newcommand{\F}{{\mathbb F}}
\newcommand{\Tr}{\mathrm{Tr}}
\newtheorem{theorem}{Theorem}
\newtheorem{definition}{Definition}
\newtheorem{lemma}{Lemma}

\newtheorem{corollary}{Corollary}

\newtheorem{algorithm}{Algorithm}

\newenvironment{proof}
  {\noindent \textbf{Proof:}}
  {\qed}

\def\qed{\ifmmode\square\else{\unskip\nobreak\hfil
\penalty50\hskip1em\null\nobreak\hfil$\square$
\parfillskip=0pt\finalhyphendemerits=0\endgraf}\fi}

\newcommand{\ket}[1]{|#1\rangle}

\newcommand{\distD}{ {\mathcal D} }
\newcommand{\om}[2]{\omega_{#1}^{#2}}

\newcommand{\smfrac}[2]{\mbox{$\frac{#1}{#2}$}}
\newcommand{\leg}[2]{\left( \frac{#1}{#2} \right)}
\newcommand{\smleg}[2]{\left({\frac{#1}{#2}}\right)}
\newcommand{\e}{\mathrm{e}}
\renewcommand{\i}{\mathrm{i}}
\newcommand{\ignore}[1]{}

\title{\textbf{Efficient Quantum Algorithms 
for Shifted \\ Quadratic Character Problems}}
\author{Wim van Dam \\
UC Berkeley, CWI Amsterdam\\
vandam@cs.berkeley.edu
\and
Sean Hallgren \\
MSRI \\
hallgren@cs.berkeley.edu}
\begin{document}
\maketitle

\begin{abstract}
  We introduce the Shifted Legendre Symbol Problem and some variants
  along with efficient quantum algorithms to solve them.  The problems
  and their algorithms are different from previous work on quantum 
  computation in that they do not appear to fit into the framework of 
  the Hidden Subgroup Problem.  
  The classical complexity of the problem is unknown, 
  despite the various results on the irregularity of Legendre sequences.
\end{abstract}

\section{Introduction}

All known problems that have a polynomial time quantum algorithms but
have no known polynomial time classical algorithm are some variant of
the Hidden Subgroup Problem.  The problem is: given a function on a
group $G$ that is constant and distinct on cosets of some unknown
subgroup, find a set of generators of the subgroup.  An example of a
problem that can be viewed in this framework is Shor's algorithm for
factoring.  In this case the problem reduces to finding the period of
a function, which amounts to finding the hidden subgroup of a cyclic
group.  The variant in this case is that the group size is unknown.
Much of the research in quantum algorithms has focused on first
reducing the problem to a variant of the Hidden Subgroup Problem
(\HSP), and then extending the machinery to handle the particular
variant.  Some examples of this include \cite{BonehL1995,
  EttingerH1998, GSV,STOC::HalesH00, STOC::HallgrenRT2000,
  Kitaev1995,MosEke98,SICOMP::Shor1997,SICOMP::Simon1997:1474}.

In this paper we introduce the Shifted Legendre Symbol Problem
(\SLSP), which does not appear to be an instance of the Hidden
Subgroup Problem.  As such, the quantum algorithm for the \SLSP
deviates from the structure of the algorithms that solve the \HSP.
The quantum component of the previous algorithms has two basic steps.
The first is to set up an equal superposition over group elements by
computing the Fourier transform, after which some function evaluation
is executed, such as $\ket{x} \longrightarrow \ket{x,f(x)}$ or
$\ket{x} \longrightarrow (-1)^{f(x)} \ket{x}$.  The second step is to
Fourier sample the state \cite{SICOMP::BernsteinV1997}, i.e. to
compute the Fourier transform and measure.  The algorithm in this
article starts with the state setup as before, but Fourier sampling is
not sufficient, because after the second step the resulting
distribution is uniform no matter what instance of the problem is
given.  This is very different from the \HSP algorithm, where the
distribution induced by Fourier sampling is enough to solve the
problem.  There is also another difference.  In general, a Fourier
transform is defined in terms of a group, and as a result previous
algorithms relied solely on properties of groups, such as what their
irreducible representations are.  One variant of the \SLSP we solve is
its extension to general finite fields.  In contrast to previous
algorithms, our algorithm uses a transform that depends on the fact
that there is an underlying field, which supports both addition and
multiplication.

The Shifted Legendre Symbol Problem is defined as follows.  Given a
function $f$ and an odd prime $p$ such that $f(x)=\smleg{x+s}{p}$,
find $s$.  Here $\smleg{x}{p}$ is the Legendre Symbol, which is $1$ if
$x$ is a square mod $p$, $-1$ if it is not, and $0$ if $p|x$.  We also
introduce a few variants of this problem.  The first variation is the
Shifted Jacobi Symbol Problem.  The setup is the same as the original
problem, except that instead of a prime $p$, an odd square free
$n=p_1\cdots p_k$ is used and the function is $f_s(x)=\smleg{x+s}{n}$,
where $\smleg{x}{n}$ is the Jacobi symbol.  The second variant also
keeps $n$ unknown.  To still be able to define the function a new
domain of size $M$ is used where $M$ is an arbitrary integer much
larger than $n$ and the shifted Jacobi symbol is repeated out to $M$.
The quantum algorithm uses a property of the Jacobi symbol to first
find the period $n$ of the function, and then uses the algorithm for
the Shifted Jacobi Symbol Problem.  The last variant is a
generalization of the \SLSP to general fields.  Here the function is a
shifted version of the quadratic character $\chi$ over the finite
field: $\chi(x)$ is $1$ if $x$ is a square in $\F_q$, $-1$ if it is
not a square, and $\chi(0)=0$.  The algorithm in this case is more
complicated and involves properties not existing in purely group
theoretic problems.  In particular the trace of field elements is used
in the transform.

The classical complexities of the problems are unknown.  When the
field size is known, the problem has polynomial query complexity, but
it is unknown if they have polynomial time algorithms.
%Proving an oracle lower bound on the classical complexity may be
%difficult because the distribution of quadratic residues is not well
%understood.  
In any case, we hope that the general structure of our algorithm will
lead to new quantum algorithms.

{\bf Related Work.}  Finding an efficient quantum algorithm for the
Shifted Legendre Symbol Problem was originally posed as an open
question in van Dam \cite{weighingmatrix}.

Many papers have studied the properties of Legendre and Jacobi
sequences, as referenced in \cite{gausssum,Damgard, Peralta}.  In
cryptography, Damg{\aa}rd \cite{Damgard} has suggested using shifted
Legendre and Jacobi sequences as pseudo-random bits in the following
sense.  The seed to the generators are two unknown values $s$ and $p$.
Consider the sequence $\smleg{s}{p}, \smleg{s+1}{p}, \ldots,
\smleg{s+t-1}{p}$, where $t$ is a polynomial in $\log p$.  If it is
hard to predict the next bits $\smleg{s+t}{p},\smleg{s+t+1}{p},\ldots$
from this sequence, then we consider the Legendre sequence
`unpredictable'.  Damg{\aa}rd showed that if Legendre sequences are
unpredictable in a very weak sense, then Jacobi sequences (defined
similarly) are unpredictable in a very strong sense.  Whether or not
Legendre sequences are indeed unpredictable is still an open question.

The \SLSP with unknown $p$ is at least as hard as the problem posed by
Damg{\aa}rd in the sense that solving the \SLSP yields the value $s$,
with which the next bits can be computed.  However, it is also
potentially easier to solve because an \SLSP algorithm is allowed to
query the string adaptively.

\ignore{
Peralta \cite{Peralta} examined the distributions of the Legendre
sequences.  A corollary related to the problem of proving an oracle
lower bound for the \SLSP is: for a fixed string $\in\{-1,+1\}^t$, 
the number of occurrences of that string in 
$\smleg{1}{p},\smleg{2}{p}, \ldots, \smleg{p-1}{p}$ is in the range 
$\frac{p}{2^t} \pm t(3+\sqrt{p})$.  
So for the \SLSP, even if we assume that an algorithm
 queries consecutive locations,
since it will pick more than $t=\log p$ values, the above bound cannot
be used to claim that the sequence occurs only once.
}

Current problems that have exponential separations between their
quantum and classical complexity can be viewed as variants of the
Hidden Subgroup Problem.  The algorithm to solve these problems first
creates a state that is uniform over a coset of a subgroup and then
computes the Fourier transform and measures.  The recursive Fourier
sampling problem \cite{SICOMP::BernsteinV1997} is not directly an
instance, but uses the same exact properties in reverse.  That is, in
the Hidden Subgroup Problem algorithm the Fourier transform takes a
state that is uniform over a coset to a perp subgroup state where the
coset is encoded in the phases of the basis vectors.  The reverse
operation is to start with the phases, and to compute the Fourier
transform to find the coset.  One level of the recursive Fourier
sampling problem does this with the subgroup restricted to trivial:
the map is $\sum_{x\in \Z_2^n} (-1)^{x\cdot s} \ket{x} \longrightarrow
\ket{s}$, where $s$ is the coset.  This problem is perhaps the closest
in structure to ours, but as mentioned, the new algorithms presented
here do not rely only on this property of cosets being taken to perp
subgroups with phases.

\section{Preliminaries}

In this section we first define the problems we solve and then we
provide other background necessary for the rest of the paper.

For a prime $p$, the {\em Legendre Symbol} $\smleg{x}{p}$ is defined to
be $1$ if $x$ is a quadratic residue, $-1$ if $x$ is a quadratic
non-residue modulo $p$, and $0$ if $p|x$.  The Legendre symbol can be
extended in several ways.  Here we will do so by defining it for for
rings $\Z_n$ and finite fields $\F_q$.  For an integer $n=p_1\cdots
p_k$ the {\em Jacobi Symbol} $\smleg{x}{n}$ is defined by
$\smleg{x}{n}=\smleg{x}{p_1}\cdots\smleg{x}{p_k}$, where the respective
$\smleg{x}{p_i}$ are Legendre Symbols and the product is over all the
prime factors $p_i$ of $n$, with repetitions.  For a finite field
$\F_q$ and $x\in \F_q$, the {\em quadratic character} $\chi(x)$ is $1$
if $x$ is a quadratic residue, $-1$ is $x$ is a quadratic non-residue,
and $0$ if $x=0$.

We can now define the problems solved in this paper.  The first
problem is the basic example which the others build on.

\begin{definition}[Shifted Legendre Symbol Problem]  
  Given an odd prime $p$ and a function $f_s:\F_p \rightarrow
  \{-1,0,1\}$ such that $f_s(x)=\smleg{x+s}{p}$ for some $s\in \F_p$,
  find $s$.
\end{definition}

The first variant extends the definition to rings.

\begin{definition}[Shifted Jacobi Symbol Problem]
  Given a square free odd integer $n$ and a function $f_s:\Z_n
  \rightarrow \{-1,0,1\}$ such that $f_s(x) = \smleg{x+s}{n}$.  Find the
  unknown shift factor $s\in\Z_n$.
\end{definition}
If the integer $n$ is not square free, the Shifted Jacobi Problem does
not have a unique answer anymore.  
Consider for example the equality, 
$\smleg{x+p}{p^2}=\smleg{x+p}{p}\smleg{x+p}{p} = \smleg{x}{p}\smleg{x}{p}=
\smleg{x}{p^2}$, for all $x\in\Z_{p^2}$.
Instead we could define the task to find one of the values $s'$ such
that $f_s(x)=\smleg{x+s'}{n}$. This problem is again efficiently
solvable on a quantum computer.

The goal of the second variant is to also keep $n$ unknown in the
Shifted Jacobi Symbol Problem.  Notice that the \SLSP with $p$ unknown
is a special case of this problem.

\begin{definition}[Shifted Jacobi Symbol Problem, unknown $\boldsymbol{n}$]
  Given an integer $M$ and a function $f_s:\Z_M \rightarrow
  \{-1,0,1\}$ such that $f_s(x)= \smleg{x+s}{n}$ for some integer 
  odd square free $n$,
  with $n^2 < M$, find $s$ and $n$.
\end{definition}

The last variant is a generalization to all possible finite fields.

\begin{definition}[Shifted Quadratic Character Problem]
  Given $q=p^r$, a power of an odd prime $p$, and a function $f: \F_q
  \rightarrow \{-1,0,1\}$ such that $f(x)=\chi(x+s)$ for some $s\in
  \F_q$, find $s$. Here $\chi$ is the quadratic character of $\F_q$.
\end{definition}

We will now give some background on finite fields, the representation
we use, and the time it takes to do basic computations.

Consider the finite field $\F_q$ with $q$ the $r$th power of the prime
$p$.  In this article the elements $x\in\F_q$ are represented as
polynomials in $\F_p[X]$ modulo an irreducible polynomial in $\F_p[X]$
of degree $r$.  The $\F_p$ coefficients of such a polynomial are
denoted by $x_j$, that is $x = \sum_{j=0}^{r-1}{x_j X^{j}}$.  When we
write $\ket{x}$ we mean $\ket{x_0,x_1,\ldots,x_{r-1}}$.  With this
representation the bit-complexity of adding or subtracting two
elements of $\F_q$ is $O(\log q)$.  Multiplication and division
require $O((\log q)^2)$ bit operations for this `model' for $\F_q$.
(See, for example, Chapter~6 in \cite{BachShallit} for details on
this.)

For a finite field $\F_{p^r}$ the \emph{trace} of an element $x$ is
defined by
\begin{eqnarray*}
  \Tr(x) & = & \sum_{j=0}^{r-1}{x^{p^j}}.
\end{eqnarray*}
By the equality $(\Tr(x))^p = \Tr(x)$ we see that the trace maps the
elements of the finite field to its base field $\F_p$. Because
$\Tr(x)$ is a polynomial of degree $p^{r-1}$ (less than $p^r$), it is
a non-constant function.  We also have, using $(x+y)^p=x^p+y^p$,
\begin{eqnarray*}
  \Tr(ax+by) & = & a\Tr(x)+b\Tr(y)
\end{eqnarray*}
for all $a,b \in \F_p$ and $x,y\in\F_q$.  It follows that the trace of
$x = \sum_{j=0}^{r-1}{x_jX^j}$ equals the summation
$\sum_{j=0}^{r-1}{x_j\Tr(X^j)}$ over the base field $\F_p$.  With this
property it can be shown that the calculation of $\Tr(x)$ requires
$O((\log q)^2)$ bit operations \cite{BachShallit}.

The main operation used in quantum algorithms is the Fourier
transform.  Here we will need to compute the Fourier transform over
$\Z_p$ for a large prime $p$, which is defined by
\begin{eqnarray*}
  \ket{x} & \longrightarrow & 
  \frac{1}{\sqrt{p}}\sum_{y=0}^{p-1}{\e^{2\pi \i(xy)/p}\ket{y}}.
\end{eqnarray*}
It is unknown how to efficiently compute this transformation exactly.
Approximations have been given in \cite{STOC::HalesH00,Kitaev1995}.
From \cite{STOC::HalesH00} we have: there is a quantum algorithm which
$\epsilon$-approximates the quantum Fourier transform over $\Z_p$ for
an arbitrary $n$-bit $p$ and any $\epsilon$ and which runs in time
$O\left(n\log\frac{n}{\epsilon}+\log^2\frac{1}{\epsilon}\right)$.  We
will denote the $p$th root of unity $\e^{2\pi \i/p}$ by
$\om{p}{}$.

We will also need a result about Fourier sampling (computing the
Fourier transform and measuring) repeated superpositions
\cite{STOC::HalesH00}.  Suppose we want to compare the distribution
induced by Fourier sampling a state $\ket{\phi}$ with the distribution
induced by Fourier sampling the state $\ket{\tilde{\phi}}$ which is
$\ket{\phi}$ repeated many times.  It turns out this is possible but
we need to define special distributions to do it.  The problem is that
$\ket{\tilde{\phi}}$ has a larger support, so we need a way to shrink
the domain so the two can be compared.  The way to do it is to use
continued fractions on the result of the sample.  We will now
formalize this.

\newcommand{\subphi}{{\scriptscriptstyle {\ket{\phi}}}}
\newcommand{\subtphi}{{\scriptscriptstyle {\ket{\tilde{\phi}}}}}

For simplicity we will suppress a detail or two to make this more
readable.  Let $\ket{\phi} = \sum_{x=0}^{n-1}{\phi_x \ket{x}}$ be an
arbitrary superposition, and let $\hat{\distD}_\subphi$ the
distribution induced by Fourier sampling $\ket{\phi}$.  Let the
superposition $\ket{\tilde{\phi}} = c\cdot 
\sum_{x=0}^{M-1}{\phi_{x \bmod n} \ket{x}}$ be $\ket{\phi}$ repeated until 
some arbitrary integer $M$, not necessarily a multiple of $n$, 
where $c$ is the proper normalization constant.  
Let $\hat{\distD}_\subtphi$ be the
distribution induced by Fourier sampling $\ket{\tilde{\phi}}$.  Notice
that $\hat{\distD}_\subphi$ is a distribution on $\{0,\ldots, n-1\}$
and $\hat{\distD}_\subtphi$ is a distribution on $\{0,\ldots, M-1\}$.

\newcommand{\rf}{{\tiny \mbox{RF}}}
\newcommand{\cf}{{\tiny \mbox{CF}}}

We can now define the two distributions we will compare.  Let
$\hat{\distD}_\subphi^\rf$ be the distribution induced on the reduced
fractions of $\hat{\distD}_\subphi$, that is, if $x$ is a sample from
$\hat{\distD}_\subphi$, we will return the fraction $x/n$ in lowest
terms.  In particular, define
$\hat{\distD}_\subphi^\rf(j,k)=\hat{\distD}_\subphi(jm)$ if $mk=n$.
Let $\hat{\distD}_\subtphi^\cf$ be the distribution induced on
fractions from first sampling $\hat{\distD}_\subtphi$ and then running
continued fractions on the result and $M$.  If $M =
\Omega(\frac{n}{\epsilon^2})$ and $M =\Omega(\frac{M}{\epsilon})$,
then $|\hat{\distD}_\subphi^\rf - \hat{\distD}_\subtphi^\cf |_1$ is
upper bounded by about $n/\sqrt{M}$.

This basically says that to understand the distribution induced by
Fourier sampling a repeated state, only the distribution induced by
Fourier sampling non-repeated state has to be understood.  However, it
is not the exact distribution of the unrepeated state, since we look
at the distribution over reduced fractions.  We will use this to solve
the unknown $n$ case of the Jacobi problem.

\section{An Algorithm for Prime Size Fields}

In this section we give algorithms solving the Shifted Legendre Symbol
Problem and variants when working over a finite field of prime size.
The main ideas are contained in the algorithm for the Shifted Legendre
Symbol Problem, and we can apply the same algorithm to solve the same
problem when $p$ is unknown, and also to solve the Shifted Jacobi
Symbol Problem.

The idea for the algorithm follows from a few known facts.  Assume we
start the algorithm in the standard way, i.e. by putting the function
value in the phase to get $\ket{f_s} = \sum_{i\in \Z_p} \leg{i+s}{p}
\ket{i}$.  Assume the functions $f_i$ are orthogonal (they are close
to orthogonal).  Define the matrix $C$ where the $i^{th}$ row is
$\ket{f_i}$.  Our quantum state $\ket{f_s}$ is one of the rows, so $C
\ket{f_s} = \ket{s}$.  The issue now is how to efficiently implement
$C$.  $C$ is a circulant matrix, i.e. $c_{i,j} = c_{i+1,j+1}$.  The
Fourier transform diagonalizes a circulant matrix: $C = F_p (F_p^{-1}
C F_p ) F_p^{-1} = F_p D F_p^{-1}$, where $D$ is diagonal, so we can
implement $C$ if we can implement $D$.  It turns out that the vector
on the diagonal of $D$ is the vector $F_p \ket{f_0}$, but $\ket{f_0}$
is an eigenvector of the Fourier transform, so up to a global phase
which we can ignore, we are done.  To summarize: to implement $C$, we
compute the Fourier transform, compute $f_0$ into the phases (this is
just the Legendre Symbol), and then compute the Fourier transform
again (it is not important whether we use $F_p$ or $F_p^{-1}$).  We
will now present this algorithm step-by-step.

\begin{algorithm}[Shifted Legendre Symbol Problem] \label{alg:p-known}
  \ \linebreak {\bf Input:} An odd prime $p$ and a function
  $f_s$ such that  $f_s(x)=\smleg{x+s}{p}$ for all $x\in\Z_p$. \\
  {\bf Output:} $s$.

  \begin{enumerate}
  \item Compute the Fourier transform over $\Z_p$ of $\ket{0}$ and
    compute $f_s$ into the phases, approximating:
    $$\frac{1}{\sqrt{p-1}} \sum_{x\in \F_p} \leg{x+s}{p} \ket{x}$$
  \item Compute the Fourier transform over $\Z_p$:
    $$\frac{1}{\sqrt{p-1}} \sum_{y\in \F_p} \om{p}{-ys} \leg{y}{p}
    \ket{y} $$
  \item Compute $f_0$ into the phases, approximating:
    $$\frac{1}{\sqrt{p}} \sum_{y\in \F_p} \om{p}{-ys} \ket{y}$$
  \item Compute the inverse Fourier transform over $\Z_p$; this gives
    the answer $\ket{-s}$.
  \end{enumerate}
\end{algorithm}

\begin{theorem} 
  Algorithm~\ref{alg:p-known} solves the Shifted Legendre Symbol
  Problem in two queries and polynomial time with probability
  exponentially close to one.
\end{theorem}

\begin{proof}
  The first step is a standard setup used in quantum algorithms.  The
  only difference is that $f_s$ evaluates to zero in one position.  In
  this case, just treat it as a one.  After this the state is
  exponentially close to the state shown.  Recall that the Legendre
  Symbol $\smleg{x}{p}$ is zero when $p|x$, so one amplitude is zero.
  
  The result of applying the Fourier transform is (where we replace
  $x$ with $x-s$)

  \begin{eqnarray*}
    \frac{1}{\sqrt{p-1}} \sum_{x=0}^{p-1} \leg{x+s}{p} \ket{x}
    &\longrightarrow& 
    \frac{1}{\sqrt{p-1}} \sum_{y=0}^{p-1}
    \frac{1}{\sqrt{p}} \sum_{x=0}^{p-1} \leg{x}{p}\om{p}{y(x-s)}
    \ket{y}.
  \end{eqnarray*}
  
  Factoring out the $\om{p}{-ys}$ term, using the change of variable
  $z=xy$, and using the facts that $\smleg{zy^{-1}}{p} = \smleg{z}{p}
  \smleg{y^{-1}}{p}$ and $\smleg{y^{-1}}{p} = \smleg{y}{p}$ we have
  $$\frac{1}{\sqrt{p-1}} \frac{1}{\sqrt{p}} \left[{\sum_{z=0}^{p-1}
      \leg{z}{p}\om{p}{z}}\right] \sum_{y=1}^{p-1} \leg{y}{p}
  \om{p}{-ys} \ket{y}$$
  So we are left to evaluate $\sum_{z=0}^{p-1}
  \smleg{z}{p}\om{p}{z}$, which is the Gau{\ss} sum
  $\cite{gausssum,Terras}$, and is $\sqrt{p}$ if $p \equiv 1 \bmod 4$
  and is $\i\sqrt{p}$ if $p \equiv 3 \bmod 4$.  Hence, up to a
  global constant which we can ignore, the state follows.
\end{proof}

\begin{corollary} 
  \label{thm:sjsp}
  Algorithm~\ref{alg:p-known} can be used to solve the Shifted Jacobi
  Symbol Problem.
\end{corollary}

\begin{proof}  
  We start with the uniform superposition of $\Z_n$ and calculate the
  function value $f_s$ for each element:
  \begin{eqnarray*}
    \frac{1}{\sqrt{n}}{\sum_{x \in \Z_n}\ket{x,0}} 
    & \longrightarrow & 
    \frac{1}{\sqrt{n}} \sum_{x \in \Z_n} \ket{x,{\textstyle \leg{x+s}{n}}}. 
  \end{eqnarray*}
  Next, we measure if the rightmost value is non-zero.  If this is the
  case, which happens with probability $\phi(n)/n$ (where $\phi$ is
  Euler's phi function obeying $\phi(n)=|\Z^*_n|$), the state has
  collapsed to the superposition:
  $$
  \frac{1}{\sqrt{\phi(n)}} \sum_{x\in\Z_n^*} \ket{x,{\textstyle
      \leg{x+s}{n}}}.
  $$
  Otherwise, we simply try again the same procedure.  (The success
  probability $\phi(n)/n$ is lower bounded by $\Omega(1/\log(\log n))$, 
  see \cite{BachShallit}, hence we can expect to be successful
  after $O(\log(\log n))$ trials.)
  
  We continue with the reduced state by changing the phase of
  $\ket{x}$ to $\smleg{x+s}{n}$ and uncomputing the function value
  again, giving
  $$
  \frac{1}{\sqrt{\phi(n)}}\sum_{x\in\Z_n}{\leg{x+s}{n}\ket{x}}.
  $$
  Let $n=p_1\cdot p_2\cdots p_k$ be the prime decomposition of $n$
  such that $\Z_n=\Z_{p_1} \times\cdots \times \Z_{p_k}$.  
  Using Shor's algorithm\cite{SICOMP::Shor1997}, we 
  can determine these factors efficiently.
  Because
  $\smleg{x+s}{n}= \smleg{x+s_1}{p_1}\cdot \smleg{x+s_2}{p_2} \cdots \smleg{x+s_k}{p_k}$, 
  we can just consider each $p_j$ component separately (with $s_1 \equiv
  s \bmod p_1$, $s_2 \equiv s \bmod p_2$, et cetera).  Hence, by
  performing the `inverse Chinese remainder' map $\ket{x}
  \longrightarrow \ket{x \bmod p_1,\ldots, x \bmod p_k}$, we obtain
  the state
  \begin{eqnarray*}
    \sum_{x_1\in\Z_{p_1}} \cdots \sum_{x_k\in\Z_{p_k}} 
    {\leg{x_1+s_1}{p_1} \cdots \leg{x_k+s_k}{p_k}}
    \ket{x_1,\ldots,x_k} 
    & = &  
    \bigotimes_{j=1}^{k}{\sum_{x_j\in\Z_{p_j}} \leg{x_j+s_j}{p_j} \ket{x_j}}.
  \end{eqnarray*}
  But now we use Algorithm~\ref{alg:p-known} on each factor to get
  $\ket{-s_1,\ldots,-s_k}$, after which the Chinese remainder theorem
  gives us the answer $s$.
\end{proof}

We now give an algorithm for the above problem when also $n$ is unknown.  
In addition to using known techniques, the algorithm depends on the fact
that sampling the Fourier transform of the shifted Legendre Symbol
results in the uniform distribution on $\Z_n^*$.

\begin{algorithm} [Shifted Jacobi Symbol Problem, unknown $\boldsymbol{n}$] 
  \label{alg:p-unknown}
  \ \linebreak {\bf Input:} An integer $M$ and a function
  $f_s:\{0,\ldots,M-1\}\rightarrow\{-1,0,1\}$ such that
  $f_s(x)= \smleg{x+s}{n}$ for some integer $n$, with $n^2 < M$ \\
  {\bf Output:} $n$ and $s$.

  \begin{enumerate}
  \item Create the following state as in Corollary~\ref{thm:sjsp}:
    $$c\cdot \sum_{x=0}^{M-1} \leg{x+s}{n} \ket{x}$$
  \item Compute the Fourier transform over $\Z_M$.
  \item\label{alg:cf} Measure, with outcome $i$, and use continued
    fractions on $i$ and $M$, returning $j/n$.
  \item Run Algorithm~\ref{alg:p-known} using $f_s$ and $n$.
  \end{enumerate}
\end{algorithm}

\begin{theorem} 
  Algorithm~\ref{alg:p-known} solves the Shifted Jacobi Symbol Problem
  with unknown $n$ in quantum polynomial time with high probability.
\end{theorem}

\begin{proof}
  Let $\ket{\psi_s} = \frac{1}{\sqrt{\phi(n)}} \sum_{x=0}^{n-1}
  \smleg{x+s}{n} \ket{x}$ be the state after the setup in
  Corollary~\ref{thm:sjsp} and let $\ket{\tilde{\psi}_s} = c
  \sum_{x=0}^{M-1} \smleg{x+s}{n}\ket{x}$ be the repeated version in
  Algorithm~\ref{alg:p-unknown}, where $c$ is the normalizing
  constant.  We can relate the distributions induced by Fourier
  sampling $\ket{\phi_s}$ and $\ket{\tilde{\phi}_s}$ using the
  discussion in Section~2.  If $M=n$ then Lemma~\ref{thm:jac} implies
  that $i$ is uniformly distributed over $\Z_n^*$ and we would be done
  since the denominator returned by continued fractions is $n$ in this
  case.  However this will still be the case even if $M\neq n$.  If
  $M$ is a multiple of $n$ and if the Fourier transform of
  $\ket{\psi_s}$ is $\sum_{x=0}^{n-1} \alpha_x \ket{x}$, then the
  Fourier transform of $\ket{\tilde{\psi}_s}$ is $\sum_{x=0}^{n-1}
  \alpha_x \ket{M/n\cdot x}$, so we get what we want.  If $M$ is not a
  multiple but is large enough, the distributions as discussed in
  Section~2 are $\epsilon$-close.
\end{proof}

\begin{lemma}
  \label{thm:jac}
Let $n$ be an odd square free integer.
If we apply the quantum Fourier transform over $\Z_n$ to the
superposition of the states $\smleg{x+s}{n}\ket{x}$ for all
$x\in\Z_n$, we establish the evolution
\begin{eqnarray*}
\frac{1}{\sqrt{\phi(n)}}\sum_{x \in \Z_n}{\leg{x+s}{n}\ket{x}}
& \longrightarrow & 
\frac{
\i^{(n-1)^2/4}
}{\sqrt{\phi(n)}}
\sum_{y \in \Z_n}{\om{n}{-sy}\leg{y}{n}\ket{y}}. 
\end{eqnarray*} 
\end{lemma}
\begin{proof}
First, we note that we can rewrite the output as 
\begin{eqnarray*}
\frac{1}{\sqrt{n\cdot \phi(n)}}
\sum_{y \in \Z_n}{{\sum_{x \in \Z_n}{\leg{x+s}{n}\om{n}{xy}}}\ket{y}} 
& = & 
\frac{1}{\sqrt{n\cdot \phi(n)}}
\sum_{y\in\Z_n}{\om{n}{-sy}
\left[{\sum_{x \in \Z_n^*}{\leg{x}{n}\om{n}{xy}}}\right]\ket{y}}, 
\end{eqnarray*}
by substituting $x$ with $x+s$ in the summation and using 
the fact that $\smleg{x}{n}=0$ for all $x \not\in \Z^*_n$.

The amplitudes between the square brackets depend on $y$ in the following way.
First, we consider the case when $y$ is co-prime to $n$, 
that is: $y \in \Z_n^*$, and there exists also an inverse $y^{-1}\in\Z^*_n$.
We then see that 
\begin{eqnarray*}
\sum_{x \in \Z_n^*}{\leg{x}{n}\om{n}{xy}} 
& = & 
\leg{y^{-1}}{n}\sum_{z \in \Z_n^*}{\leg{z}{n}\om{n}{z}},
\end{eqnarray*}
where we used the substitution $x \leftarrow zy^{-1}$ and the multiplicativity
of the Jacobi symbol. 

Next, we look at the case where $n$ and $y$ have a common, non-trivial,
factor $f$. We say that $n=mf$ and $y=rf$, and we know that $f$ and $m$ 
are co-prime (because $n$ is square free). 
The Chinese remainder theorem tells us that there is a 
bijection between the elements $x\in\Z_n$ and the 
coordinates $(x\bmod{m},x\bmod{f}) \in \Z_m \times \Z_f$,
which also establishes a one-to-one mapping between 
$\Z^*_n$ and $\Z_m^* \times \Z_f^*$.
This allows us to rewrite the expression as follows.
\begin{eqnarray*}
\sum_{x \in \Z_n^*}{\leg{x}{n}\om{n}{xy}} 
& = & 
\sum_{x \in \Z_{mf}^*}{\leg{x}{mf}\om{mf}{xrf}} \\
& = & 
\sum_{x \in \Z_{mf}^*}{\leg{x\bmod{m}}{m}\leg{x\bmod{f}}{f}\om{m}{xr}} \\
& = & 
\sum_{x_1\in\Z_m^*}{\leg{x_1}{m}\om{m}{x_1r}\sum_{x_2\in \Z_{f}^*}{\leg{x_2}{f}}}. 
\end{eqnarray*}
Because $f$ is odd and square free $\sum_{x\in\Z_f^*}{\smleg{x}{f}}=0$,
and hence the above term equals zero.
This concludes the proof of the lemma.
\end{proof}

\section{An Algorithm for General Finite Fields}

Here we will solve the general case of the Shifted Legendre Symbol
Problem for any finite field $\F_q$.  From now on $q=p^r$, with $p$ an
odd prime and the degree $r$ an integer.  See Section~2 for details
about finite fields.  The idea used for the SLSP algorithm cannot be
used directly here, since the matrix is no longer circulant.  To get
around that problem, we use the following map:

\begin{lemma}[Trace-Fourier Transform over $\F_q$] \label{thm:TFT}
  The unitary mapping 
  \begin{eqnarray*}
    \ket{x}& \longrightarrow & 
    \frac{1}{\sqrt{q}} \sum_{y\in \F_q}  \om{p}{\Tr(xy)} \ket{y}
  \end{eqnarray*} 
  is computable in polynomial time.
\end{lemma}

\begin{proof}
  Assume that the mapping
  \begin{eqnarray*}
    \ket{x} & \longrightarrow & \bigotimes_{j=0}^{r-1}{\ket{\Tr(xX^j)}}. 
  \end{eqnarray*}
  can be computed in polynomial time.  First apply this map, and then
  compute the Fourier transform over $\Z_p^r$.  This gives us the
  final state
  \begin{eqnarray*}
    \bigotimes_{j=0}^{r-1} \frac{1}{\sqrt{p}} \sum_{y_j\in\F_p} 
    \om{p}{\Tr(xX^j)y_j}\ket{y_j}
    & = & \frac{1}{\sqrt{q}} \sum_{y\in \F_q}  \om{p}{\Tr(xy)} \ket{y}.
\end{eqnarray*}

We will now show that the map
\begin{eqnarray*}
  \ket{x} & \longrightarrow & \ket{\Tr(x),\Tr(xX),\ldots,\Tr(xX^{r-1})}
\end{eqnarray*}
is reversible.  Let $T(x)= [\Tr(x), \Tr(xX), \ldots, \Tr(xX^{r-1})]$.
$T$ is a linear functional since $\Tr$ is, so if $T(a)=T(b)$ then
$T(a-b)$ is the zero vector.  We will show that $T(x)$ is not the zero
vector except for $x=0$.  Suppose $T(x)$ is the zero vector.  Since
$\Tr$ is not the zero map, choose $a \in \F_q$ such that $\Tr(a) \neq
0$.  Choose $z_0, \ldots, z_{r-1}$ such that $\sum_j z_j x X^j = a$.
Then $\Tr(a) = \Tr( \sum_j z_j x X^j ) = \sum_j z_j \Tr( x X^j) = 0$,
since $\Tr(x X^j)=0$ for all $j$.  But this is a contraction by the
choice of $a$.  So $T$ is one-to-one.

We will now show that the map is computable in polynomial time.  It is
enough if $x$ can be computed from $\Tr(x), \Tr(xX), \ldots,
\Tr(xX^{r-1})$.  But the equations $\Tr(x) = \sum_{j=0}^{r-1} x_j
\Tr(X^j)$, $\Tr(xX) = \sum_{j=0}^{r-1} x_j \Tr(X^{j+1})$, \ldots,
$\Tr(xX^{r-1}) = \sum_{j=0}^{r-1} x_j \Tr(X^{j+r-1})$ are $r$ linear
equations in $r$ unknowns, and the values $\Tr(x), \Tr(xX), \ldots,
\Tr(xX^{r-1})$ and $\Tr(X^j)$ for all $j$ are known, so the
coefficients $x_j$ of $x$ can be solved for using linear algebra.
\end{proof}
(We recently learned that, independently, de Beaudrap \emph{et al.\ }
\cite{Beaudrapetal} have used a transform closely related to the above
for the construction of a different quantum algorithm.)

\begin{theorem} Algorithm~\ref{alg:general-fields} (see below) solves the 
  Shifted Quadratic Character Problem over any finite field with two
  queries and in polynomial time with probability exponentially close
  to one.
\end{theorem}

The presentation of the algorithm below differs from the SLSP
algorithm in that it does not use approximations.  Approximations work
here also, but here we show how the problem can be solved exactly if
the base field is of fixed size (so that the Fourier transform is not
approximate).  Also notice that the Trace-Fourier transform is not a
unique solution to this problem, any linear functional will work in
place of $\Tr$.

\begin{algorithm} \label{alg:general-fields}
  \ \linebreak {\bf Input:} A power of a prime $q=p^r$ and a function
  $f_s$ such that $f_s(x)=\chi(x+s)$. \\
  {\bf Output:} $s$.

  \begin{enumerate}
  \item Use the Fourier transform over $\Z_{q+1}$ on $\ket{0}$, and
    two queries to $f_s$ to create (with probability $q/(q+1)$) the
    state
    $$\frac{1}{\sqrt{q}} \sum_{x\in \F_q} \chi(x+s) \ket{x} +
    \frac{1}{\sqrt{q}}\ket{\delta}.$$
    With probability $1/(q+1)$ this
    step gives $s$ directly.
  \item Compute the Trace-Fourier transform of Lemma~\ref{thm:TFT}
    over $\F_q$.  In the proof it will be shown that the output of
    this transform equals
    \begin{eqnarray}\label{eq:step2}
      & &  \frac{1}{q}\left[{\sum_{z\in\F_q}{\chi(z)\om{p}{\Tr(z)}}}\right]
      \left({\sum_{y\in\F_q}{\chi(y)\om{p}{\Tr(-sy)}\ket{y}}}\right) + 
      \frac{1}{\sqrt{q}}\ket{\delta}. 
    \end{eqnarray}
    The term between the square brackets is known the \emph{quadratic
      Gau\ss{} sum} $G(\F_q)$, with $G(\F_q) =
%    (-1)^{r-1}\i^{r(p-1)^2/4} \sqrt{q}$ (see the appendix of this
%    article and Theorem~11.5.4 in \cite{gausssum}).  With this
    (-1)^{r-1}\i^{r(p-1)^2/4} \sqrt{q}$ (see Theorem~11.5.4 in
    \cite{gausssum}).  With this knowledge we can perform the next
    step.
  \item Uncompute the phases $\chi(y)$ for $y\neq 0$, and change the
    dummy vector to $(-1)^{r-1}\i^{r(p-1)^2/4}\ket{0}$.  By dropping
    the general phase $G(\F_q)/\sqrt{q}$, we can now write
    $$\frac{1}{\sqrt{q}} \sum_{y\in \F_q} \om{p}{\Tr(-sy)}
    \ket{y}$$
    for the state.
  \item Finally, compute the inverse Trace-Fourier transform over
    $\F_q$.  This gives us the requested shift parameter as
    $\ket{-s}$.
  \end{enumerate} 
\end{algorithm} 

\begin{proof}
  For the first step we create, with one call to $f_s$ the
  superposition
  $$\frac{1}{\sqrt{q+1}}\sum_{x\in\F_q}{\ket{x,f_s(x)}} \quad +\quad
  \frac{1}{\sqrt{q+1}}\ket{\delta,1},$$
  where $\delta$ denotes a
  `dummy state'.  Next, we measure if the rightmost bit is zero. If
  this is the case (probability $1/(q+1)$), the state has collapsed to
  $\ket{-s,0}$, which tells us the value $s$ immediately.  Otherwise,
  we are left with the superposition of the entries $x$ with
  $f_s(x)=\pm 1$ and the dummy state. This enables us to create the
  proper phases $f_s(x)=\chi(x+s)$ and uncompute (with a second $f_s$
  query) the rightmost bit, which we will ignore from now on.
  
  At step 2, we perform the Trace-Fourier transform to the state,
  yielding
  $$ 
  \frac{1}{q}\sum_{x\in\F_q}{\sum_{y \in
      \F_q}{\chi(x+s)\om{p}{\Tr(xy)}\ket{y}}}
  \quad + \quad \frac{1}{\sqrt{q}}\ket{\delta}.
  $$
  We rewrite this expression as follows: replace $x$ with
  $z=xy+sy$, and use the multiplicativity of $\chi$ and the linearity
  of the trace in $\chi(zy^{-1})\om{p}{\Tr(z-sy)} =
  \chi(z)\om{p}{\Tr(z)}\cdot\chi(y)\om{p}{\Tr(-sy)}$.  This proves the
  validity of Equation~\ref{eq:step2}.
\end{proof}

\section{Conclusion and Open Problems}
We have shown the existence of efficient quantum algorithms for
several versions of the `Shifted Quadratic Character Problem'.  The
classical complexity of these problems remains open.  In the light of
Shor's result\cite{SICOMP::Shor1997}, we would also like to know
whether the problems become classically tractable if we assume that
factoring is easy.

\section{Acknowledgements}
WvD was supported by the Institute for Logic, Language 
and Computation in Amsterdam, the EU fifth framework project 
QAIP IST-1999-11234, and the TALENT grant S 62-552 of the
Netherlands Organization for Scientific Research (NWO).

\appendix
\section{From Legendre Symbols to Gau{\ss} Sums}
\subsection{Legendre Symbol, Jacobi Symbol, and Quadratic Character}
In this appendix, $p$ is a prime, $q$ is the prime power $p^r$ with 
degree $r$, and $n$ is a (typically non-prime) positive integer.
We denote the field of size $p$ by $\Z_p$, instead of the perhaps more 
accurate $\Z/(p\Z)$.  Similarly, the ring induced by $\bmod{n}$ 
addition and multiplication is $\Z_n$.  The finite field of size $q$ is 
described by $\F_q$, and hence $\F_p = \Z_p$, but $\F_{p^2}\neq \Z_{p^2}$.
The respective multiplicative subgroups are indicated by the $*$ superscript:
$\Z_p^*$, $\Z_n^*$ and $\F_q^*$.

The \emph{Legendre symbol} indicates if a non-zero element 
is a square modulo $p$ or not:
\begin{eqnarray*}
\leg{x}{p} & = & \left\{
\begin{array}{rl}
0 & \mbox{if $x=0\bmod{p}$} \\
+1 & \mbox{if there exists a $y\neq 0$ such that $y^2=x \bmod{p}$}\\
-1 & \mbox{if for all $y$: $y^2\neq x \bmod{p}$.}
\end{array}
\right.
\end{eqnarray*}
Let $n=p_1\cdot p_2 \cdots p_k$ be the prime factor decomposition of $n$.  
The \emph{Jacobi symbol} generalizes the Legendre symbol for all
rings $\Z_n$ in the following way:
\begin{eqnarray*}
\leg{x}{n} & = & \leg{x}{p_1}\cdot \leg{x}{p_2}\cdots\leg{x}{p_k}.
\end{eqnarray*}
Clearly, $\smleg{x}{n}=0$ for all $x$ not co-prime to $n$. 
Note that $\smleg{x}{n}=+1$ does not always imply 
that there is a $y$ with $y^2 = x \bmod{n}$. 
Take, for example, $\smleg{2}{9}=1$.

For finite fields $\F_q$, the Legendre symbol becomes the
\emph{quadratic character} $\chi$, which is defined 
\begin{eqnarray*}
\chi(x) & = & 
\left\{
\begin{array}{rl}
0 & \mbox{if $x=0$} \\
+1 & \mbox{if there exists a $y\neq 0$ such that $y^2=x$}\\
-1 & \mbox{if for all $y$: $y^2\neq x$.}
\end{array}
\right.
\end{eqnarray*}
for all $x\in\F_q$.
 
\subsection{Basic Properties}
The Legendre symbol, Jacobi symbol and quadratic character are 
all three \emph{multiplicative characters} because they obey
$(xy/p)=(x/p)(y/p)$, 
$(xy/n)=(x/n)(y/n)$, and
$\chi(xy)=\chi(x)\chi(y)$, respectively.
This implies a series of results.

Let $g$ be a generator of $\Z_p^*$.  Because the 
multiplicative subgroup has $p-1$ elements, we know
that $g^i = g^j\bmod{p}$ if and only if $i=j\bmod{p-1}$.
Hence, the quadratic equation $(g^j)^2 = g^{2j}=g^i\bmod{p}$ 
is correct if and only if $2j = i \bmod{p-1}$. 

For an odd prime $p$, there can only exists a $j$ 
with $(g^j)^2 = g^i\bmod{p}$ when $i$ is even, 
as $p-1$ is even.
Obviously, if $i$ is even, then
$g^j$ with $j=\smfrac{i}{2}$ gives also a solution.
In short: $(g^i/p) = (-1)^i$.
 This proves that $\smfrac{p-1}{2}$ of the elements
$x$ of $\Z_p^*$ are a quadratic residue with 
$(x/p)=+1$, while the other $\smfrac{p-1}{2}$
are non-squares.

If $p$ is even, then for all $i$ either $i$ or $(i+p-1)$ 
will be even. Hence, either $j=\smfrac{i}{2}$ or
$j=\smfrac{i+p-1}{2}$ gives a proper solution to the
equality $(g^j)^2=g^i\bmod{p}$.
This proves that all elements  $x\in\Z^*_p$ are 
quadratic residues with $(x/p)=+1$.
This rather redundant proof for the only existing case
$p=2$ is justified by the following lemma.
\begin{lemma}
For any finite field $\F_{p^r}$, we have for the summation
of its quadratic character values
\begin{eqnarray*}
\sum_{x\in\F_{p^r}}{\chi(x)} & = & 
\left\{
\begin{array}{rl}
0 & \mbox{if $p$ is odd} \\
p^r-1 & \mbox{if $p$ is even.}
\end{array}
\right.
\end{eqnarray*}
\end{lemma}
\begin{proof}
Every multiplicative group $\F^*_{p^r}$ has a generator $g$
with period $p^r-1$.
Use this in combination with the proof method of the preceding 
paragraphs.
\end{proof}

We will reach a similar result for the summation of the 
Jacobi symbol values over $\Z_n$, when $n$ is odd and squarefree.
Let again $n=p_1\cdots p_k$ be the prime decomposition.
The Chinese remainder theorem tells us that the mapping 
$x \in\Z_n \rightarrow 
(x\bmod{p_1},\ldots,x\bmod{p_k}) \in \Z_{p_1} \times \cdots \times \Z_{p_k}$ 
is a bijection.
(All $p_i$ terms are different, because we assumed $n$ to be square free.) 
This enables us to prove the following lemma.
\begin{lemma}
Let $n$ be an odd, square free integer.  The summation of all 
the Jacobi values of $\Z_n$ obeys
\begin{eqnarray*}
\sum_{x\in\Z_n}{\leg{x}{n}} & = & 0.
\end{eqnarray*}
\end{lemma}
\begin{proof}
Let $n=p_1\cdots p_k$ be the decomposition of $n$ into 
its prime factors.
The definition of the Jacobi symbol in combination with
Chinese remainder theorem yields the equality
\begin{eqnarray*}
\sum_{x\in\Z_n}{\leg{x}{n}} & = & 
\sum_{x\in\Z_n}{\leg{x}{p_1}\cdots\leg{x}{p_k}} \\
& = & 
\sum_{x_1\in\Z_{p_1}}{\cdots 
\sum_{x_k\in\Z_{p_k}}{
\leg{x_1}{p_1}\cdots\leg{x_k}{p_k}}} \\
& = & 
\left({\sum_{x_1\in\Z_{p_1}}{\leg{x_1}{p_1}}}\right)
\cdots
\left({\sum_{x_k\in\Z_{p_k}}{\leg{x_k}{p_k}}}\right).
\end{eqnarray*}
By the previous lemma we know that each 
$\sum_{x\in\Z_p}{(x/p)}$ is zero, hence the 
above product is zero as well.
\end{proof}

\subsection{Gau{\ss} Sums}
Let $\om{p}{}$ denote the complex root $\e^{2\pi i/p}$.
The \emph{trace} of an element $x\in\F_{p^r}$ is defined by
$\Tr(x) = \sum_{j=0}^{r-1}{x^{p^j}}$.
It can be shown that for every $x\in\F_{p^r}$, its trace is 
an element of the base-field: $\Tr(x)\in\F_p$.
When we write $\om{p}{\Tr(x)}$ we interpret the value
$\Tr(x)$ as an element of the set $\{0,1,\ldots,p-1\} \subset \Z$. 
\begin{definition}
For the field $\Z_p$, the ring $\Z_n$ and the finite field $\F_{p^r}$
we define the \emph{quadratic Gau{\ss} sum} $G$ by
\begin{eqnarray*}
G(\Z_p) & = & \sum_{x\in\Z_p}{\leg{x}{p}\om{p}{x}}, \\
G(\Z_n) & = & \sum_{x\in\Z_n}{\leg{x}{n}\om{n}{x}}, \\
G(\F_{p^r}) & = & \sum_{x\in\F_{p^r}}{\chi(x)\om{p}{\Tr(x)}}.
\end{eqnarray*}
\end{definition}
It is not immediately clear that this definition does not 
give contradicting values for the identical cases 
$G(\Z_p)$ and $G(\F_p)$. 
However, the next result shows that this conflict does
not occur. We will not give the proofs of the following
lemma as that goes far beyond the scope of this article.
Instead, the curious reader is referred to the book by 
Berndt \emph{et al.}\cite{gausssum}

\begin{lemma}
Let $p$ be an odd prime and $n$ an odd square free integer. 
The following equalities hold for the different 
quadratic Gau{\ss} sums:
\begin{eqnarray*}
G(\Z_p) & = &  
\left\{\begin{array}{rl}
\sqrt{p} & \mbox{if $p=1\bmod{4}$}\\
\i\sqrt{p} & \mbox{if $p=3\bmod{4}$}
\end{array}
\right.\\
G(\Z_n) & = &
\left\{\begin{array}{rl}
\sqrt{n} & \mbox{if $n=1\bmod{4}$}\\
\i\sqrt{n} & \mbox{if $n=3\bmod{4}$}
\end{array}
\right.\\  
G(\F_{p^r}) & = &
\left\{\begin{array}{rl}
-\sqrt{p^r} & \mbox{if $p=1\bmod{4}$ and $r$ is even} \\
 \sqrt{p^r} & \mbox{if $p=1\bmod{4}$ and $r$ is odd} \\
-\sqrt{p^r} & \mbox{if $p=3\bmod{4}$ and $r=0\bmod{4}$} \\
\i\sqrt{p^r} & \mbox{if $p=3\bmod{4}$ and $r=1\bmod{4}$} \\
 \sqrt{p^r} & \mbox{if $p=3\bmod{4}$ and $r=2\bmod{4}$} \\
-\i\sqrt{p^r} & \mbox{if $p=3\bmod{4}$ and $r=3\bmod{4}$} \\
\end{array}
\right.  
\end{eqnarray*}
\end{lemma}
Note that indeed $G(\Z_p)=G(\F_p)$.

\end{document}